\documentclass[10pt,aps,prl,twocolumn,preprintnumbers,superscriptaddress]{revtex4-1}

\usepackage{verbatim}
\usepackage{amsmath}
\usepackage{amsfonts}
\usepackage{amssymb}
\usepackage{bbm}
\usepackage{latexsym}
\usepackage{hhline}
\usepackage{graphicx}

\usepackage{hyperref} 

\usepackage{etex}
\usepackage[DIV12]{typearea}
\usepackage{amsmath,amsfonts,amssymb}
\usepackage{graphicx}
\usepackage{mathrsfs}
\usepackage{bbm}
\usepackage{dsfont}
\usepackage{pifont}
\usepackage{units}
\usepackage{longtable}
\usepackage[small,loose]{subfigure}  
\usepackage{cancel}
\usepackage{booktabs}
\usepackage{braket}
\usepackage{mathtools}
\usepackage{pdfcomment}

\usepackage{color}
\definecolor{darkgreen}{rgb}{0.0, 0.7, 0.0}

\newcommand{\Eqref}[1]{Equation~\eqref{#1}}
\newcommand{\Figref}[1]{Figure~\ref{#1}}

\newcommand{\SU}[1]{\ensuremath{\mathrm{SU}(#1)}}

\newcommand{\U}[1]{\ensuremath{\mathrm{U}(#1)}}
\newcommand{\Z}[1]{\ensuremath{\mathbbm{Z}_{#1}}} 

\newcommand{\rep}[1]{\ensuremath{\boldsymbol{#1}}}
\newcommand{\crep}[1]{\ensuremath{\overline{\boldsymbol{#1}}}}

\newcommand{\Id}{\ensuremath{\mathbbm{1}}}

\DeclareMathOperator{\diag}{diag}

\begin{document}

%
%
\title{Grand Unification without Proton Decay}

\date{\today}
\author{Andreas \surname{M\"utter}}
\email[]{andreas.muetter@ph.tum.de}
\affiliation{Physik Department T30, Technische Universit\"at M\"unchen, James--Franck--Stra\ss e, 85748 Garching, Germany}
\author{Michael \surname{Ratz}}
\email[]{michael.ratz@tum.de}
\affiliation{Physik Department T30, Technische Universit\"at M\"unchen, James--Franck--Stra\ss e, 85748 Garching, Germany}
\author{Patrick K.S. \surname{Vaudrevange}}
\email[]{patrick.vaudrevange@tum.de}
\affiliation{Physik Department T30, Technische Universit\"at M\"unchen, James--Franck--Stra\ss e, 85748 Garching, Germany}
\affiliation{Excellence Cluster Universe, Technische Universit\"at M\"unchen, Boltzmannstr. 2, D-85748, Garching, Germany}
\begin{abstract}
It is commonly believed that grand unified theories (GUTs) predict proton
decay. This is because the exchange of extra GUT gauge bosons gives rise to
dimension 6 proton decay operators. We show that there exists a class of GUTs in
which these operators are absent. Many string and supergravity models in the
literature belong to this class.
\end{abstract}
%
\preprint{TUM-HEP 1046/16; FLAVOUR-EU 123; LMU-ASC 23/16} 

\maketitle

\section{Introduction}

\begin{figure}[!h!]
\includegraphics{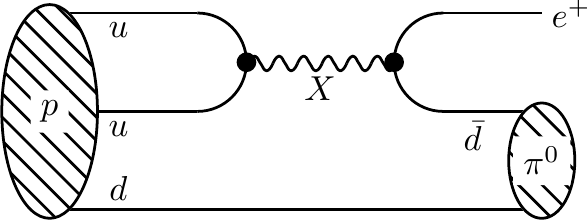}
\caption{Proton decay operator of dimension~6.}
\label{fig:ProtonDecayDimension6}
\end{figure}

Grand unified theories (GUTs) are a hypothetical framework that unifies three
out of four known forces, electromagnetism, the weak force and strong
interactions in a larger symmetry group $G_\mathrm{GUT}$. The perhaps greatest
virtue of GUTs is that they provide a compelling explanation of the structure
and quantum numbers of standard model (SM) matter. In \SU5, the lepton doublets
$\ell$ and $d$--type quarks $\bar d$ get combined in \crep{5}--plets while the
remaining three representations of a generation, i.e.\ $q$, $\bar u$ and $\bar
e$, transform as a \rep{10}--plet. It is commonly believed
that the most compelling ``smoking gun'' signature of GUTs is proton decay. This
is because GUTs are endowed with extra ``$X$'' gauge bosons from the coset
$\SU5/G_\mathrm{SM}$, where $G_\mathrm{SM}=\SU3_\mathrm{C}\times\SU2_\mathrm{L}\times
\U1_Y$ is the SM gauge group. These gauge bosons mediate transitions
between $\ell$ and $\bar d$ as well as $q$, $\bar u$ and $\bar e$, thus inducing
dimension 6 proton decay operators (cf.\
\Figref{fig:ProtonDecayDimension6}). 

\smallskip

\begin{figure}[!h!]
\includegraphics{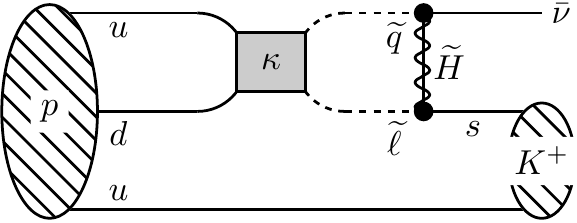}
\caption{Proton decay operator of dimension~5. Here, $\widetilde{H}$, $\widetilde{q}$ and $\widetilde{\ell}$ stand for the superpartners of the Higgs and the quark and lepton doublets.}
\label{fig:ProtonDecayDimension5}
\end{figure}

The traditional approach is to break the GUT symmetry spontaneously
\cite{Georgi:1974sy,Fritzsch:1974nn}. However, the framework of supersymmetric
grand unification (SUSY GUTs) typically suffers from too rapid proton decay (see
e.g.\ \cite{Dermisek:2000hr,Murayama:2001ur}). More specifically, the exchange
of the \SU5 partners of the electroweak Higgs doublets, the so--called Higgs
triplets, leads to dimension 5 proton decay operators
\cite{Sakai:1981pk,Dimopoulos:1981dw}. These induce the decay mode $p\to
K^+\,\bar\nu$ (see \Figref{fig:ProtonDecayDimension5}). That is, generically
SUSY GUTs predict this proton decay signature.

\smallskip

This mode can be eliminated by either the Babu--Barr mechanism
\cite{Babu:1993we} or by going to higher dimensions. Specifically, in the
so--called orbifold GUTs~\cite{Kawamura:1999nj,Kawamura:2000ev}, this mode is
automatically absent \cite{Altarelli:2001qj}. As is well known, the
dimension 4 operators can be forbidden by $R$ parity \cite{Farrar:1978xj}. One
is then left with the dimension 6 proton decay, which is induced by gauge bosons
transforming in the coset $\SU5/G_\mathrm{SM}$, and thus believed to be
\emph{the} smoking gun signature of unification. The purpose of this Letter
is to show that these dimension 6 operators are also absent in a class of grand
unified theories in which the GUT symmetry gets broken non--locally in extra 
dimensions.

\section{Higher--dimensional models of grand unification}

Higher--dimensional models of grand unification decompose into two classes,
models with local and models with non--local GUT symmetry breaking. Models with
local GUT symmetry breaking comprise the traditional orbifold compactifications
of the heterotic
string~\cite{Dixon:1985jw,Dixon:1986jc,Ibanez:1986tp,Ibanez:1987sn,Casas:1988hb,Font:1989aj,Lebedev:2006kn,Lebedev:2008un,Nilles:2014owa}
(see~\cite{Quevedo:1996sv,Bailin:1999nk,Nilles:2008gq} for reviews) as well as
orbifold GUTs. On the other hand, models with non--local GUT symmetry breaking
include Calabi--Yau compactifications 
(see~\cite{Braun:2005nv,Bouchard:2005ag,Anderson:2009mh,Anderson:2011ns,Anderson:2012yf,Nibbelink:2015vha}
for models that come very close to the SM) as well as field--theoretic
constructions~\cite{Hall:2001tn,Hebecker:2003we,Hebecker:2004ce,Trapletti:2006xv} 
and some orbifold compactification of the heterotic 
string~\cite{Donagi:2008xy,Fischer:2012qj}, which have only been explored more
recently.

\smallskip

To illustrate our main points, let us start by looking at one extra  dimension,
which is parametrized by $y \in F=[0,2\pi\, R]$. Take an \SU5 GUT  where the
gauge bosons are free to propagate in a fifth dimension. Furthermore,  we assume
$n_G$ generations of quarks and leptons transforming as 
$\crep{5}\oplus\rep{10}$. In our discussion, we focus on the matter 
$\crep{5}$--plets of the first two generations and denote them by 
\begin{align}
  \Psi_i(x,y) ~=~ \left(\begin{array}{c}\ell_i(x,y)\\\overline{d}_i(x,y)\end{array}\right)
 \quad(i=1,2)\;.
\end{align}

\subsection{Local GUT breaking}

\begin{figure}[!h!]
\includegraphics{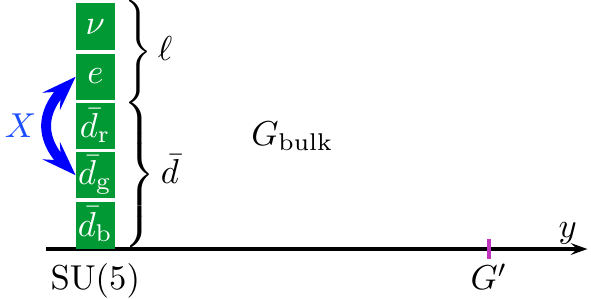}
\caption{Local \SU5 grand unification in 5D. SM multiplets are localized at
points with an \SU5 GUT symmetry. The bulk symmetry $G_\mathrm{bulk}$ gets
broken locally to \SU5 and $G'$, and the low--energy gauge group is the
intersection of these groups in $G_\mathrm{bulk}$, i.e.\ $G_\mathrm{SM}=\SU5\cap
G'$ (cf.\ \cite{Buchmuller:2005sh}). The profile of the massive $X$ bosons is
non--trivial at the points where the SM matter lives.}
\label{fig:LocalBreaking}
\end{figure}

We first discuss models with ``local breaking'', where $n_G=3$, using the
example of an $\mathbbm{S}^1/(\Z2\times\Z2)$ orbifold
\cite{Kawamura:1999nj,Kawamura:2000ev}.  Here the SM matter is localized at
fixed points with a GUT symmetry (see \Figref{fig:LocalBreaking}). Therefore,
the fields have to appear in complete GUT multiplets. The GUT symmetry gets
broken at the other fixed point, and the zero modes of the $X$ bosons get
projected out. Therefore, the profile of the $X$ bosons is non--trivial at the
points at which matter is localized. As a consequence, the Kaluza--Klein modes
of the extra gauge bosons $X \subset  \SU5/G_\mathrm{SM}$ mediate between
$d$--type quarks and lepton doublets within each $\Psi_i(x,y)$ field. That is,
there are effective interactions of the form
\begin{align}\label{eq:LintX}
 \mathscr{L}_\mathrm{eff}&~\supset~\int\!\mathrm{d}
 y\,g_{5\mathrm{D}}\,\sum_{i=1}^{n_G}
 \overline{\ell}_i(x,y)\,\gamma^\mu\,X_\mu(x,y)\,\overline{d}_i(x,y) \;,
\end{align}
where $g_{5D}$ denotes the 5D gauge coupling and $\gamma^\mu$ the 4D 
$\gamma$--matrices. Here the $X$ bosons have a mass of the order of 
$M_\mathrm{GUT}$.

\smallskip

In summary, in models with local GUT symmetry breaking the structure of SM
matter gets explained by a local GUT symmetry. Like conventional
four--dimensional GUT models, these constructions predict dimension 6 proton
decay $p\to e^+\,\pi^0$. However, unlike in most of the conventional SUSY GUTs,
here the dimension 5 proton decay mode is absent. 

\subsection{Non--local GUT breaking}\label{sec:NonLocalBreaking}

\begin{figure}[!h!]
\includegraphics{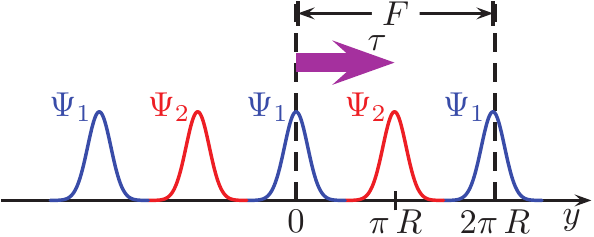}
\caption{Matter fields in the fifth dimension parametrized by $y$. Under the
action of $\tau$, $\Psi_1$ and $\Psi_2$ get identified.}
\label{fig:MatterIn5D}
\end{figure}

Let us now switch to settings in which the \SU5 symmetry gets broken
non--locally. To illustrate our points, matter fields are now assumed to be
localized in the fifth dimension, see \Figref{fig:MatterIn5D}. Later, when  we
present a stringy completion, we will discuss an orbifold compactification of
the heterotic string, where the corresponding states are localized at some fixed
planes. We start in the ``upstairs picture'' with a setting exhibiting $n_G = 6$
generations of quarks and leptons, and focus on two generations of matter
$\crep{5}$--plets $\Psi_{i}(x,y)$ for $i=1,2$.

\smallskip

In the next step, we break the \SU5 GUT group non--locally to $G_\mathrm{SM}$ by
a so--called freely--acting \Z2 with associated Wilson line. In our example,
the freely--acting \Z2 acts as a translation that, from a 5D point of view,
identifies points in the $y$--direction which differ by $\tau = \pi\, R$, i.e.\
$y \sim y + \tau$. 

In addition to the geometrical \Z2 action, due to the presence of the Wilson
line, the \SU5 gauge bosons $A^\mu(x,y)$ are subject to the non--trivial 
boundary condition
\begin{align}\label{eq:NonLocalBC}
A^\mu(x,y) ~\mapsto~ A^\mu(x,y + \tau) ~\stackrel{!}{=}~ P\, A^\mu(x,y)\, P^{-1}\;.
\end{align}
Here $A^\mu(x,y) = A_a^\mu(x,y)\, \mathsf{T}_a$ with \SU5 generators
$\mathsf{T}_a$ and
\begin{align}\label{eq:P}
 P ~=~ \diag(-1,-1,+1,+1,+1)
\end{align}
with $P^2 = \Id$. This \Z2 boundary condition projects out the zero modes of the
extra gauge bosons $X \subset \SU5/G_\mathrm{SM}$, and, hence, breaks
\SU5 to $G_\mathrm{SM}$. 

In addition, the freely--acting translation by $\tau$ identifies the two 
fields $\Psi_{1}(x,y)$ and $\Psi_{2}(x,y)$. We therefore obtain a
non--trivial \Z2 boundary condition for the matter $\crep{5}$--plets
$\Psi_i(x,y)$, i.e.\
\begin{align}
  \Psi_1(x,y)~\xmapsto{\tau}~ \Psi_1(x,y+\tau) ~\stackrel{!}{=}~ P\, \Psi_2(x,y) \;.
\end{align}
Then, the $d$--type quark and the lepton doublet of the first SM generation are 
given by the \Z2 invariant linear combinations
\begin{subequations}\label{eq:LinearCombinations}
\begin{align}
 \ell(x,y)        &~=~\frac{1}{\sqrt{2}}\left[\ell_1(x,y) - \ell_2(x,y)\right]\;, \\
 \overline{d}(x,y)&~=~\frac{1}{\sqrt{2}}\left[\overline{d}_1(x,y) + \overline{d}_2(x,y)\right]\;.
\end{align}
\end{subequations}
The orthogonal field directions
\begin{subequations}\label{eq:OrthogonalLinearCombinations}
\begin{align}
 \ell^{(\perp)}(x,y)        &~=~\frac{1}{\sqrt{2}}\left[\ell_1(x,y) + \ell_2(x,y)\right]\;, \\
 \overline{d}^{(\perp)}(x,y)&~=~\frac{1}{\sqrt{2}}\left[\overline{d}_1(x,y) - \overline{d}_2(x,y)\right]\;,
\end{align}
\end{subequations}
are projected out in 4D. Thus, two ``upstairs'' generations of matter 
$\crep{5}$--plets are combined to the $d$--type quark and the lepton doublet of 
the first SM generation. Repeating these steps for $n_G = 6$ generations of 
$\crep{5}\oplus\rep{10}$ yields the SM with three generations. In particular, 
the matter still furnishes complete \SU5 representations!

The interactions of the first generation's $d$--type quark and the lepton doublet 
with the $X$ boson, \Eqref{eq:LintX}, reads in the new field basis
\begin{align}
 \mathscr{L}_\mathrm{eff} 
  &~\supset~ \int\!\mathrm{d} y\,g_{5\mathrm{D}}\,
  \left[ \overline{\ell}(x,y)\,\gamma^\mu\,X_\mu(x,y)\,\overline{d}^{(\perp)}(x,y) 
  \right.\nonumber\\
  &\quad{} +\left.\, \overline{\ell}^{(\perp)}(x,y)\,\gamma^\mu\,X_\mu(x,y)\,\overline{d}(x,y)
 \right]\;.
\end{align}
There is no interaction of the physical $d$ quark and the lepton doublet, 
i.e.\ $\mathscr{L}_\mathrm{eff} \not\supset \overline{\ell}\,\gamma^\mu\,X_\mu\,\overline{d}$. 
Hence, the dimension 6 proton decay operator, which usually arises from 
integrating out the $X$ bosons in \Figref{fig:ProtonDecayDimension6}, does not 
appear.

\smallskip

The absence of the dimension 6 proton decay operator can also be understood in 
terms of a $\Z2^\tau$ symmetry that acts on quarks, leptons and the extra gauge bosons 
as
\begin{subequations}
\begin{align}
 \ell(x,y)         &~\xmapsto{~\Z2^\tau~}~ + \ell(x,y) \;,\\
 \overline{d}(x,y) &~\xmapsto{~\Z2^\tau~}~ + \overline{d}(x,y)\;,\\
 X^\mu(x,y)        &~\xmapsto{~\Z2^\tau~}~ - X^\mu(x,y)\;.
\end{align}
\end{subequations}
However, this symmetry does not imply that the profiles of the $X$ bosons
vanish in the regions where SM matter lives.

\section{A stringy completion}\label{sec:StringyCompletion}

Let us now study a string--derived setup that realizes the scenario of
non--locally broken GUTs discussed above. We consider a
$\mathbbm{T}^6/(\Z2\times\Z2)$ orbifold compactification of the
heterotic string as discussed in~\cite{Blaszczyk:2009in}. A more detailed
description of the model will be presented elsewhere \cite{Muetter:2016pr}. The
actual orbifold model has six compact dimensions, but it is sufficient to study
three of the extra dimensions $\vec y=(y_2,y_4,y_6)^T \in\mathbbm{R}^3$ in order
to understand the non--local breaking. 

\smallskip

The $\Z2\times\Z2$ twists $\theta$ and $\omega$ act as
\begin{subequations}\label{eq:Twists}
\begin{align}
\left(y_2,y_4,y_6\right)~\xmapsto{~\theta~}~\left( y_2,-y_4,-y_6\right)\;,\\
\left(y_2,y_4,y_6\right)~\xmapsto{~\omega~}~\left(-y_2, y_4,-y_6\right)\;.
\end{align}
\end{subequations}
In a first step, these twists act on an orthogonal three--torus $\mathbbm{T}^3$
spanned by
\begin{subequations}\label{eq:LatticeVectors}
\begin{align}
\vec e_2~=~ \left(2\pi\, R_2,0,0\right)^T\;, \\
\vec e_4~=~ \left(0,2\pi\, R_4,0\right)^T\;, \\
\vec e_6~=~ \left(0,0,2\pi\, R_6\right)^T\;.
\end{align}
\end{subequations}
In a second step, we include a freely--acting translation 
\begin{align}\label{eq:TauFreely}
\vec\tau ~=~\frac{1}{2}\,\left(\vec e_2+\vec e_4+\vec e_6\right)\;,
\end{align}
which renders $\mathbbm{T}^3$ non--factorizable, and yields a non--trivial 
fundamental group $\pi_1 = \Z2$ for the resulting orbifold.
\smallskip

The model is constructed in such a way that prior to the action of $\vec\tau$
the \SU5 gauge bosons $A^\mu$ survive the orbifold projection. This amounts to
requiring the boundary conditions
\begin{subequations}\label{eq:BCondA}
\begin{align}
A^\mu(x,\vec y+ n_i\,\vec e_i)        ~=~A^\mu(x,\vec y)\;,\label{eq:BCondNormalTransl}\\
A^\mu(x,\theta^k\,\omega^\ell\,\vec y)~=~A^\mu(x,\vec y)\;,\label{eq:BCondTwist}
\end{align}
\end{subequations}
where $n_i\in\mathbbm{Z}$ and $k,\,\ell\in \left\lbrace 0,\,1\right\rbrace$. 
Next, \SU5 is broken to $G_\text{SM}$ by the freely--acting translation. In
order to achieve this, we choose a gauge embedding of $\vec\tau$, i.e.\ the
Wilson line, such that the boundary condition for the extra gauge bosons $X
\subset \SU5/G_\mathrm{SM}$ reads
\begin{align}
X^\mu(x,\vec y+\vec\tau)~=~-X^\mu(x,\vec y)\;,\label{eq:BCondFreely}
\end{align}
cf.~\Eqref{eq:NonLocalBC}. In particular, this removes the zero modes of the 
$X$ bosons.
\smallskip

The combined action of lattice translations \eqref{eq:LatticeVectors} and 
twists \eqref{eq:Twists} leaves planes fixed. These fixed planes are
determined by
\begin{align}
\theta^k\,\omega^\ell\,\vec y_\text{f}+n_i\,\vec e_i~=~\vec y_\text{f}\;.
\end{align} 
There are twelve solutions of this equation,
\begin{subequations}
\begin{align}
k=0\;,~\ell=1\quad \text{and} \quad n_2\;,~n_6\in\left\lbrace0,1\right\rbrace\;,\\
k=1\;,~\ell=0\quad \text{and} \quad n_4\;,~n_6\in\left\lbrace0,1\right\rbrace\;,\\
k=1\;,~\ell=1\quad \text{and} \quad n_2\;,~n_4\in\left\lbrace0,1\right\rbrace\;,
\end{align}
\end{subequations}
which we label by $\vec y_\text{f}{}^{(\!\alpha\!)}, \alpha=1,\dots,12$. On these planes,
one finds six generations of localized SM matter sitting in \SU5 multiplets. The
$X$ boson wave function, which satisfies \eqref{eq:BCondA} and
\eqref{eq:BCondFreely}, does not vanish on the fixed planes where the matter is
located. 
Therefore, one may na\"{i}vely expect that interaction terms of the form of
\eqref{eq:LintX} induce gauge--mediated proton decay.

\smallskip

However, a universal feature of models with non--local GUT breaking is that
there are $3n$ copies of matter in the upstairs picture. These get identified by
the freely acting symmetry of order $n$, thus reducing the number of generations
to three. Specifically, in our construction, there are three pairs of matter
fields sitting on three pairs of distinct fixed planes. Under the action of  the
freely--acting translation $\vec\tau$, the fixed plane  $\vec
y_\text{f}{}^{(\!\alpha\!)}$ gets mapped to $\vec y_\text{f}{}^{(\!\beta\!)}$,
\begin{align}
\vec y_\text{f}{}^{(\!\alpha\!)}~\xmapsto{~\vec\tau~}~\vec y_\text{f}{}^{(\!\beta\!)}~
\Leftrightarrow~\vec y_\text{f}{}^{(\!\beta\!)}~=~\vec y_\text{f}{}^{(\!\alpha\!)}
+\vec\tau\;.
\end{align}
Thus, $\vec y_\text{f}{}^{(\!\alpha\!)}$ and $\vec y_\text{f}{}^{(\!\beta\!)}$ 
get identified, and, analogous to \eqref{eq:LinearCombinations} only the 
$\vec\tau$--invariant linear combinations of the fields localized at  $\vec
y_\text{f}{}^{(\!\alpha\!)}$ and $\vec y_\text{f}{}^{(\!\beta\!)}$  survive the
projection conditions. This reduces the number of SM generations  from six to
three. The Wilson line is such that it reproduces \eqref{eq:P} for  the \SU5
representations. Hence, the $\vec\tau$--invariant linear combinations  for
quarks and leptons come either with $+$ or $-$, 
cf.~\eqref{eq:LinearCombinations}. Consequently, as in the scenario discussed 
in the 5D toy--example, the dimension 6 proton decay operators do not get
induced.

\section{Summary}
\label{sec:Summary}

We have discussed grand unified models in which the GUT symmetry gets broken
non--locally, and found that there the dimension 6 proton decay operators are
absent. Nevertheless, these settings do explain the structure of matter, i.e.\
the SM fermions are guaranteed to appear in complete GUT multiplets. That is to
say, the constructions exhibit the main virtues of grand unified theories but
do not predict proton decay, which was believed to be the smoking gun signal of
grand unification.

\smallskip

As we have mentioned, many existing string and supergravity models belong to
this class. In other words, the absence of proton decay does not concern some
exotic type of constructions, but some of the most promising string
compactifications known to date. Turning this around, one could say that proton
decay experiments can give us invaluable insights on how the SM is completed in
the ultraviolet. If the mode $p\to K^+\,\bar\nu$ was observed, this would point
towards four--dimensional SUSY GUTs. On the other hand, if this mode gets even
further constrained, one is led to higher--dimensional models of grand
unification, or non--supersymmetric GUTs. If one would see the decay $p\to
e^+\,\pi^0$ with a rate of about $1/(10^{35}\,\mathrm{y})$, this would strongly
favor settings in which the GUT symmetry gets broken locally. However, if one
does not observe this decay, this would point towards models of non--local grand
unification.

\subsection*{Acknowledgments}

We would like to thank Mu--Chun Chen, Stefan Groot Nibbelink and Andreas
Trautner for useful discussions. M.R.\ would like to thank the Aspen Center for
Physics and UC Irvine, where part of this work was done, for hospitality. This
work was partially supported by the DFG cluster of excellence ``Origin and
Structure of the Universe'' (\texttt{www.universe-cluster.de}). This research
was done in the context of the ERC Advanced Grant project ``FLAVOUR''~(267104).

\bibliographystyle{aipnum4-1}
\bibliography{Orbifold}

\begin{thebibliography}{38}%
\makeatletter
\providecommand \@ifxundefined [1]{%
 \@ifx{#1\undefined}
}%
\providecommand \@ifnum [1]{%
 \ifnum #1\expandafter \@firstoftwo
 \else \expandafter \@secondoftwo
 \fi
}%
\providecommand \@ifx [1]{%
 \ifx #1\expandafter \@firstoftwo
 \else \expandafter \@secondoftwo
 \fi
}%
\providecommand \natexlab [1]{#1}%
\providecommand \enquote  [1]{``#1''}%
\providecommand \bibnamefont  [1]{#1}%
\providecommand \bibfnamefont [1]{#1}%
\providecommand \citenamefont [1]{#1}%
\providecommand \href@noop [0]{\@secondoftwo}%
\providecommand \href [0]{\begingroup \@sanitize@url \@href}%
\providecommand \@href[1]{\@@startlink{#1}\@@href}%
\providecommand \@@href[1]{\endgroup#1\@@endlink}%
\providecommand \@sanitize@url [0]{\catcode `\\12\catcode `\$12\catcode
  `\&12\catcode `\#12\catcode `\^12\catcode `\_12\catcode `\%12\relax}%
\providecommand \@@startlink[1]{}%
\providecommand \@@endlink[0]{}%
\providecommand \url  [0]{\begingroup\@sanitize@url \@url }%
\providecommand \@url [1]{\endgroup\@href {#1}{\urlprefix }}%
\providecommand \urlprefix  [0]{URL }%
\providecommand \Eprint [0]{\href }%
\providecommand \doibase [0]{http://dx.doi.org/}%
\providecommand \selectlanguage [0]{\@gobble}%
\providecommand \bibinfo  [0]{\@secondoftwo}%
\providecommand \bibfield  [0]{\@secondoftwo}%
\providecommand \translation [1]{[#1]}%
\providecommand \BibitemOpen [0]{}%
\providecommand \bibitemStop [0]{}%
\providecommand \bibitemNoStop [0]{.\EOS\space}%
\providecommand \EOS [0]{\spacefactor3000\relax}%
\providecommand \BibitemShut  [1]{\csname bibitem#1\endcsname}%
\let\auto@bib@innerbib\@empty
\bibitem [{\citenamefont {Georgi}\ and\ \citenamefont
  {Glashow}(1974)}]{Georgi:1974sy}%
  \BibitemOpen
  \bibfield  {author} {\bibinfo {author} {\bibfnamefont {H.}~\bibnamefont
  {Georgi}}\ and\ \bibinfo {author} {\bibfnamefont {S.~L.}\ \bibnamefont
  {Glashow}},\ }\href@noop {} {\bibfield  {journal} {\bibinfo  {journal} {Phys.
  Rev. Lett.}\ }\textbf {\bibinfo {volume} {32}},\ \bibinfo {pages} {438}
  (\bibinfo {year} {1974})}\BibitemShut {NoStop}%
\bibitem [{\citenamefont {Fritzsch}\ and\ \citenamefont
  {Minkowski}(1975)}]{Fritzsch:1974nn}%
  \BibitemOpen
  \bibfield  {author} {\bibinfo {author} {\bibfnamefont {H.}~\bibnamefont
  {Fritzsch}}\ and\ \bibinfo {author} {\bibfnamefont {P.}~\bibnamefont
  {Minkowski}},\ }\href@noop {} {\bibfield  {journal} {\bibinfo  {journal}
  {Ann. Phys.}\ }\textbf {\bibinfo {volume} {93}},\ \bibinfo {pages} {193}
  (\bibinfo {year} {1975})}\BibitemShut {NoStop}%
\bibitem [{\citenamefont {Derm\'{i}\u{s}ek}, \citenamefont {Mafi},\ and\
  \citenamefont {Raby}(2001)}]{Dermisek:2000hr}%
  \BibitemOpen
  \bibfield  {author} {\bibinfo {author} {\bibfnamefont {R.}~\bibnamefont
  {Derm\'{i}\u{s}ek}}, \bibinfo {author} {\bibfnamefont {A.}~\bibnamefont
  {Mafi}}, \ and\ \bibinfo {author} {\bibfnamefont {S.}~\bibnamefont {Raby}},\
  }\href@noop {} {\bibfield  {journal} {\bibinfo  {journal} {Phys. Rev.}\
  }\textbf {\bibinfo {volume} {D63}},\ \bibinfo {pages} {035001} (\bibinfo
  {year} {2001})},\ \Eprint {http://arxiv.org/abs/hep-ph/0007213}
  {hep-ph/0007213} \BibitemShut {NoStop}%
\bibitem [{\citenamefont {Murayama}\ and\ \citenamefont
  {Pierce}(2002)}]{Murayama:2001ur}%
  \BibitemOpen
  \bibfield  {author} {\bibinfo {author} {\bibfnamefont {H.}~\bibnamefont
  {Murayama}}\ and\ \bibinfo {author} {\bibfnamefont {A.}~\bibnamefont
  {Pierce}},\ }\href {\doibase 10.1103/PhysRevD.65.055009} {\bibfield
  {journal} {\bibinfo  {journal} {Phys. Rev.}\ }\textbf {\bibinfo {volume}
  {D65}},\ \bibinfo {pages} {055009} (\bibinfo {year} {2002})},\ \Eprint
  {http://arxiv.org/abs/hep-ph/0108104} {arXiv:hep-ph/0108104 [hep-ph]}
  \BibitemShut {NoStop}%
\bibitem [{\citenamefont {Sakai}\ and\ \citenamefont
  {Yanagida}(1982)}]{Sakai:1981pk}%
  \BibitemOpen
  \bibfield  {author} {\bibinfo {author} {\bibfnamefont {N.}~\bibnamefont
  {Sakai}}\ and\ \bibinfo {author} {\bibfnamefont {T.}~\bibnamefont
  {Yanagida}},\ }\href@noop {} {\bibfield  {journal} {\bibinfo  {journal}
  {Nucl. Phys.}\ }\textbf {\bibinfo {volume} {B197}},\ \bibinfo {pages} {533}
  (\bibinfo {year} {1982})}\BibitemShut {NoStop}%
\bibitem [{\citenamefont {Dimopoulos}, \citenamefont {Raby},\ and\
  \citenamefont {Wilczek}(1982)}]{Dimopoulos:1981dw}%
  \BibitemOpen
  \bibfield  {author} {\bibinfo {author} {\bibfnamefont {S.}~\bibnamefont
  {Dimopoulos}}, \bibinfo {author} {\bibfnamefont {S.}~\bibnamefont {Raby}}, \
  and\ \bibinfo {author} {\bibfnamefont {F.}~\bibnamefont {Wilczek}},\
  }\href@noop {} {\bibfield  {journal} {\bibinfo  {journal} {Phys. Lett.}\
  }\textbf {\bibinfo {volume} {B112}},\ \bibinfo {pages} {133} (\bibinfo {year}
  {1982})}\BibitemShut {NoStop}%
\bibitem [{\citenamefont {Babu}\ and\ \citenamefont
  {Barr}(1993)}]{Babu:1993we}%
  \BibitemOpen
  \bibfield  {author} {\bibinfo {author} {\bibfnamefont {K.~S.}\ \bibnamefont
  {Babu}}\ and\ \bibinfo {author} {\bibfnamefont {S.~M.}\ \bibnamefont
  {Barr}},\ }\href@noop {} {\bibfield  {journal} {\bibinfo  {journal} {Phys.
  Rev.}\ }\textbf {\bibinfo {volume} {D48}},\ \bibinfo {pages} {5354} (\bibinfo
  {year} {1993})},\ \Eprint {http://arxiv.org/abs/hep-ph/9306242}
  {hep-ph/9306242} \BibitemShut {NoStop}%
\bibitem [{\citenamefont {Kawamura}(2000)}]{Kawamura:1999nj}%
  \BibitemOpen
  \bibfield  {author} {\bibinfo {author} {\bibfnamefont {Y.}~\bibnamefont
  {Kawamura}},\ }\href@noop {} {\bibfield  {journal} {\bibinfo  {journal}
  {Prog. Theor. Phys.}\ }\textbf {\bibinfo {volume} {103}},\ \bibinfo {pages}
  {613} (\bibinfo {year} {2000})},\ \Eprint
  {http://arxiv.org/abs/hep-ph/9902423} {hep-ph/9902423} \BibitemShut {NoStop}%
\bibitem [{\citenamefont {Kawamura}(2001)}]{Kawamura:2000ev}%
  \BibitemOpen
  \bibfield  {author} {\bibinfo {author} {\bibfnamefont {Y.}~\bibnamefont
  {Kawamura}},\ }\href@noop {} {\bibfield  {journal} {\bibinfo  {journal}
  {Prog. Theor. Phys.}\ }\textbf {\bibinfo {volume} {105}},\ \bibinfo {pages}
  {999} (\bibinfo {year} {2001})},\ \Eprint
  {http://arxiv.org/abs/hep-ph/0012125} {hep-ph/0012125} \BibitemShut {NoStop}%
\bibitem [{\citenamefont {Altarelli}\ and\ \citenamefont
  {Feruglio}(2001)}]{Altarelli:2001qj}%
  \BibitemOpen
  \bibfield  {author} {\bibinfo {author} {\bibfnamefont {G.}~\bibnamefont
  {Altarelli}}\ and\ \bibinfo {author} {\bibfnamefont {F.}~\bibnamefont
  {Feruglio}},\ }\href@noop {} {\bibfield  {journal} {\bibinfo  {journal}
  {Phys. Lett.}\ }\textbf {\bibinfo {volume} {B511}},\ \bibinfo {pages} {257}
  (\bibinfo {year} {2001})},\ \Eprint {http://arxiv.org/abs/hep-ph/0102301}
  {hep-ph/0102301} \BibitemShut {NoStop}%
\bibitem [{\citenamefont {Farrar}\ and\ \citenamefont
  {Fayet}(1978)}]{Farrar:1978xj}%
  \BibitemOpen
  \bibfield  {author} {\bibinfo {author} {\bibfnamefont {G.~R.}\ \bibnamefont
  {Farrar}}\ and\ \bibinfo {author} {\bibfnamefont {P.}~\bibnamefont {Fayet}},\
  }\href {\doibase 10.1016/0370-2693(78)90858-4} {\bibfield  {journal}
  {\bibinfo  {journal} {Phys. Lett.}\ }\textbf {\bibinfo {volume} {B76}},\
  \bibinfo {pages} {575} (\bibinfo {year} {1978})}\BibitemShut {NoStop}%
\bibitem [{\citenamefont {Dixon}\ \emph {et~al.}(1985)\citenamefont {Dixon},
  \citenamefont {Harvey}, \citenamefont {Vafa},\ and\ \citenamefont
  {Witten}}]{Dixon:1985jw}%
  \BibitemOpen
  \bibfield  {author} {\bibinfo {author} {\bibfnamefont {L.~J.}\ \bibnamefont
  {Dixon}}, \bibinfo {author} {\bibfnamefont {J.~A.}\ \bibnamefont {Harvey}},
  \bibinfo {author} {\bibfnamefont {C.}~\bibnamefont {Vafa}}, \ and\ \bibinfo
  {author} {\bibfnamefont {E.}~\bibnamefont {Witten}},\ }\href@noop {}
  {\bibfield  {journal} {\bibinfo  {journal} {Nucl. Phys.}\ }\textbf {\bibinfo
  {volume} {B261}},\ \bibinfo {pages} {678} (\bibinfo {year}
  {1985})}\BibitemShut {NoStop}%
\bibitem [{\citenamefont {Dixon}\ \emph {et~al.}(1986)\citenamefont {Dixon},
  \citenamefont {Harvey}, \citenamefont {Vafa},\ and\ \citenamefont
  {Witten}}]{Dixon:1986jc}%
  \BibitemOpen
  \bibfield  {author} {\bibinfo {author} {\bibfnamefont {L.~J.}\ \bibnamefont
  {Dixon}}, \bibinfo {author} {\bibfnamefont {J.~A.}\ \bibnamefont {Harvey}},
  \bibinfo {author} {\bibfnamefont {C.}~\bibnamefont {Vafa}}, \ and\ \bibinfo
  {author} {\bibfnamefont {E.}~\bibnamefont {Witten}},\ }\href@noop {}
  {\bibfield  {journal} {\bibinfo  {journal} {Nucl. Phys.}\ }\textbf {\bibinfo
  {volume} {B274}},\ \bibinfo {pages} {285} (\bibinfo {year}
  {1986})}\BibitemShut {NoStop}%
\bibitem [{\citenamefont {Ib{\'a}{\~n}ez}, \citenamefont {Nilles},\ and\
  \citenamefont {Quevedo}(1987)}]{Ibanez:1986tp}%
  \BibitemOpen
  \bibfield  {author} {\bibinfo {author} {\bibfnamefont {L.~E.}\ \bibnamefont
  {Ib{\'a}{\~n}ez}}, \bibinfo {author} {\bibfnamefont {H.~P.}\ \bibnamefont
  {Nilles}}, \ and\ \bibinfo {author} {\bibfnamefont {F.}~\bibnamefont
  {Quevedo}},\ }\href@noop {} {\bibfield  {journal} {\bibinfo  {journal} {Phys.
  Lett.}\ }\textbf {\bibinfo {volume} {B187}},\ \bibinfo {pages} {25} (\bibinfo
  {year} {1987})}\BibitemShut {NoStop}%
\bibitem [{\citenamefont {Ib{\'a}{\~n}ez}\ \emph {et~al.}(1987)\citenamefont
  {Ib{\'a}{\~n}ez}, \citenamefont {Kim}, \citenamefont {Nilles},\ and\
  \citenamefont {Quevedo}}]{Ibanez:1987sn}%
  \BibitemOpen
  \bibfield  {author} {\bibinfo {author} {\bibfnamefont {L.~E.}\ \bibnamefont
  {Ib{\'a}{\~n}ez}}, \bibinfo {author} {\bibfnamefont {J.~E.}\ \bibnamefont
  {Kim}}, \bibinfo {author} {\bibfnamefont {H.~P.}\ \bibnamefont {Nilles}}, \
  and\ \bibinfo {author} {\bibfnamefont {F.}~\bibnamefont {Quevedo}},\
  }\href@noop {} {\bibfield  {journal} {\bibinfo  {journal} {Phys. Lett.}\
  }\textbf {\bibinfo {volume} {B191}},\ \bibinfo {pages} {282} (\bibinfo {year}
  {1987})}\BibitemShut {NoStop}%
\bibitem [{\citenamefont {Casas}\ and\ \citenamefont
  {Mu{\~n}oz}(1988)}]{Casas:1988hb}%
  \BibitemOpen
  \bibfield  {author} {\bibinfo {author} {\bibfnamefont {J.~A.}\ \bibnamefont
  {Casas}}\ and\ \bibinfo {author} {\bibfnamefont {C.}~\bibnamefont
  {Mu{\~n}oz}},\ }\href@noop {} {\bibfield  {journal} {\bibinfo  {journal}
  {Phys. Lett.}\ }\textbf {\bibinfo {volume} {B214}},\ \bibinfo {pages} {63}
  (\bibinfo {year} {1988})}\BibitemShut {NoStop}%
\bibitem [{\citenamefont {Font}\ \emph {et~al.}(1990)\citenamefont {Font},
  \citenamefont {Ib{\'a}{\~n}ez}, \citenamefont {Quevedo},\ and\ \citenamefont
  {Sierra}}]{Font:1989aj}%
  \BibitemOpen
  \bibfield  {author} {\bibinfo {author} {\bibfnamefont {A.}~\bibnamefont
  {Font}}, \bibinfo {author} {\bibfnamefont {L.~E.}\ \bibnamefont
  {Ib{\'a}{\~n}ez}}, \bibinfo {author} {\bibfnamefont {F.}~\bibnamefont
  {Quevedo}}, \ and\ \bibinfo {author} {\bibfnamefont {A.}~\bibnamefont
  {Sierra}},\ }\href@noop {} {\bibfield  {journal} {\bibinfo  {journal} {Nucl.
  Phys.}\ }\textbf {\bibinfo {volume} {B331}},\ \bibinfo {pages} {421}
  (\bibinfo {year} {1990})}\BibitemShut {NoStop}%
\bibitem [{\citenamefont {Lebedev}\ \emph {et~al.}(2007)\citenamefont
  {Lebedev}, \citenamefont {Nilles}, \citenamefont {Raby}, \citenamefont
  {Ramos-S{\'a}nchez}, \citenamefont {Ratz}, \citenamefont {Vaudrevange},\ and\
  \citenamefont {Wingerter}}]{Lebedev:2006kn}%
  \BibitemOpen
  \bibfield  {author} {\bibinfo {author} {\bibfnamefont {O.}~\bibnamefont
  {Lebedev}}, \bibinfo {author} {\bibfnamefont {H.~P.}\ \bibnamefont {Nilles}},
  \bibinfo {author} {\bibfnamefont {S.}~\bibnamefont {Raby}}, \bibinfo {author}
  {\bibfnamefont {S.}~\bibnamefont {Ramos-S{\'a}nchez}}, \bibinfo {author}
  {\bibfnamefont {M.}~\bibnamefont {Ratz}}, \bibinfo {author} {\bibfnamefont
  {P.~K.~S.}\ \bibnamefont {Vaudrevange}}, \ and\ \bibinfo {author}
  {\bibfnamefont {A.}~\bibnamefont {Wingerter}},\ }\href@noop {} {\bibfield
  {journal} {\bibinfo  {journal} {Phys. Lett.}\ }\textbf {\bibinfo {volume}
  {B645}},\ \bibinfo {pages} {88} (\bibinfo {year} {2007})},\ \Eprint
  {http://arxiv.org/abs/hep-th/0611095} {hep-th/0611095} \BibitemShut {NoStop}%
\bibitem [{\citenamefont {Lebedev}\ \emph {et~al.}(2008)\citenamefont
  {Lebedev}, \citenamefont {Nilles}, \citenamefont {Ramos-S\'{a}nchez},
  \citenamefont {Ratz},\ and\ \citenamefont {Vaudrevange}}]{Lebedev:2008un}%
  \BibitemOpen
  \bibfield  {author} {\bibinfo {author} {\bibfnamefont {O.}~\bibnamefont
  {Lebedev}}, \bibinfo {author} {\bibfnamefont {H.~P.}\ \bibnamefont {Nilles}},
  \bibinfo {author} {\bibfnamefont {S.}~\bibnamefont {Ramos-S\'{a}nchez}},
  \bibinfo {author} {\bibfnamefont {M.}~\bibnamefont {Ratz}}, \ and\ \bibinfo
  {author} {\bibfnamefont {P.~K.~S.}\ \bibnamefont {Vaudrevange}},\ }\href
  {\doibase 10.1016/j.physletb.2008.08.054} {\bibfield  {journal} {\bibinfo
  {journal} {Phys. Lett.}\ }\textbf {\bibinfo {volume} {B668}},\ \bibinfo
  {pages} {331} (\bibinfo {year} {2008})},\ \Eprint
  {http://arxiv.org/abs/0807.4384} {arXiv:0807.4384 [hep-th]} \BibitemShut
  {NoStop}%
\bibitem [{\citenamefont {Nilles}\ and\ \citenamefont
  {Vaudrevange}(2015)}]{Nilles:2014owa}%
  \BibitemOpen
  \bibfield  {author} {\bibinfo {author} {\bibfnamefont {H.~P.}\ \bibnamefont
  {Nilles}}\ and\ \bibinfo {author} {\bibfnamefont {P.~K.~S.}\ \bibnamefont
  {Vaudrevange}},\ }\href {\doibase 10.1142/S0217732315300086} {\bibfield
  {journal} {\bibinfo  {journal} {Mod. Phys. Lett.}\ }\textbf {\bibinfo
  {volume} {A30}},\ \bibinfo {pages} {1530008} (\bibinfo {year} {2015})},\
  \Eprint {http://arxiv.org/abs/1403.1597} {arXiv:1403.1597 [hep-th]}
  \BibitemShut {NoStop}%
\bibitem [{\citenamefont {Quevedo}(1996)}]{Quevedo:1996sv}%
  \BibitemOpen
  \bibfield  {author} {\bibinfo {author} {\bibfnamefont {F.}~\bibnamefont
  {Quevedo}},\ }\href {\doibase 10.1063/1.49735} {\  (\bibinfo {year} {1996}),\
  10.1063/1.49735},\ \bibinfo {note} {[AIP Conf. Proc.359,202(1996)]},\ \Eprint
  {http://arxiv.org/abs/hep-th/9603074} {arXiv:hep-th/9603074} \BibitemShut
  {NoStop}%
\bibitem [{\citenamefont {Bailin}\ and\ \citenamefont
  {Love}(1999)}]{Bailin:1999nk}%
  \BibitemOpen
  \bibfield  {author} {\bibinfo {author} {\bibfnamefont {D.}~\bibnamefont
  {Bailin}}\ and\ \bibinfo {author} {\bibfnamefont {A.}~\bibnamefont {Love}},\
  }\href@noop {} {\bibfield  {journal} {\bibinfo  {journal} {Phys. Rept.}\
  }\textbf {\bibinfo {volume} {315}},\ \bibinfo {pages} {285} (\bibinfo {year}
  {1999})}\BibitemShut {NoStop}%
\bibitem [{\citenamefont {Nilles}\ \emph {et~al.}(2009)\citenamefont {Nilles},
  \citenamefont {Ramos-S{\'a}nchez}, \citenamefont {Ratz},\ and\ \citenamefont
  {Vaudrevange}}]{Nilles:2008gq}%
  \BibitemOpen
  \bibfield  {author} {\bibinfo {author} {\bibfnamefont {H.~P.}\ \bibnamefont
  {Nilles}}, \bibinfo {author} {\bibfnamefont {S.}~\bibnamefont
  {Ramos-S{\'a}nchez}}, \bibinfo {author} {\bibfnamefont {M.}~\bibnamefont
  {Ratz}}, \ and\ \bibinfo {author} {\bibfnamefont {P.~K.~S.}\ \bibnamefont
  {Vaudrevange}},\ }\href {\doibase 10.1140/epjc/s10052-008-0740-1} {\bibfield
  {journal} {\bibinfo  {journal} {Eur. Phys. J.}\ }\textbf {\bibinfo {volume}
  {C59}},\ \bibinfo {pages} {249} (\bibinfo {year} {2009})},\ \Eprint
  {http://arxiv.org/abs/0806.3905} {arXiv:0806.3905 [hep-th]} \BibitemShut
  {NoStop}%
\bibitem [{\citenamefont {Braun}\ \emph {et~al.}(2006)\citenamefont {Braun},
  \citenamefont {He}, \citenamefont {Ovrut},\ and\ \citenamefont
  {Pantev}}]{Braun:2005nv}%
  \BibitemOpen
  \bibfield  {author} {\bibinfo {author} {\bibfnamefont {V.}~\bibnamefont
  {Braun}}, \bibinfo {author} {\bibfnamefont {Y.-H.}\ \bibnamefont {He}},
  \bibinfo {author} {\bibfnamefont {B.~A.}\ \bibnamefont {Ovrut}}, \ and\
  \bibinfo {author} {\bibfnamefont {T.}~\bibnamefont {Pantev}},\ }\href@noop {}
  {\bibfield  {journal} {\bibinfo  {journal} {JHEP}\ }\textbf {\bibinfo
  {volume} {05}},\ \bibinfo {pages} {043} (\bibinfo {year} {2006})},\ \Eprint
  {http://arxiv.org/abs/hep-th/0512177} {hep-th/0512177} \BibitemShut {NoStop}%
\bibitem [{\citenamefont {Bouchard}\ and\ \citenamefont
  {Donagi}(2006)}]{Bouchard:2005ag}%
  \BibitemOpen
  \bibfield  {author} {\bibinfo {author} {\bibfnamefont {V.}~\bibnamefont
  {Bouchard}}\ and\ \bibinfo {author} {\bibfnamefont {R.}~\bibnamefont
  {Donagi}},\ }\href@noop {} {\bibfield  {journal} {\bibinfo  {journal} {Phys.
  Lett.}\ }\textbf {\bibinfo {volume} {B633}},\ \bibinfo {pages} {783}
  (\bibinfo {year} {2006})},\ \Eprint {http://arxiv.org/abs/hep-th/0512149}
  {hep-th/0512149} \BibitemShut {NoStop}%
\bibitem [{\citenamefont {Anderson}\ \emph {et~al.}(2010)\citenamefont
  {Anderson}, \citenamefont {Gray}, \citenamefont {He},\ and\ \citenamefont
  {Lukas}}]{Anderson:2009mh}%
  \BibitemOpen
  \bibfield  {author} {\bibinfo {author} {\bibfnamefont {L.~B.}\ \bibnamefont
  {Anderson}}, \bibinfo {author} {\bibfnamefont {J.}~\bibnamefont {Gray}},
  \bibinfo {author} {\bibfnamefont {Y.-H.}\ \bibnamefont {He}}, \ and\ \bibinfo
  {author} {\bibfnamefont {A.}~\bibnamefont {Lukas}},\ }\href {\doibase
  10.1007/JHEP02(2010)054} {\bibfield  {journal} {\bibinfo  {journal} {JHEP}\
  }\textbf {\bibinfo {volume} {1002}},\ \bibinfo {pages} {054} (\bibinfo {year}
  {2010})},\ \Eprint {http://arxiv.org/abs/0911.1569} {arXiv:0911.1569
  [hep-th]} \BibitemShut {NoStop}%
\bibitem [{\citenamefont {Anderson}\ \emph {et~al.}(2011)\citenamefont
  {Anderson}, \citenamefont {Gray}, \citenamefont {Lukas},\ and\ \citenamefont
  {Palti}}]{Anderson:2011ns}%
  \BibitemOpen
  \bibfield  {author} {\bibinfo {author} {\bibfnamefont {L.~B.}\ \bibnamefont
  {Anderson}}, \bibinfo {author} {\bibfnamefont {J.}~\bibnamefont {Gray}},
  \bibinfo {author} {\bibfnamefont {A.}~\bibnamefont {Lukas}}, \ and\ \bibinfo
  {author} {\bibfnamefont {E.}~\bibnamefont {Palti}},\ }\href {\doibase
  10.1103/PhysRevD.84.106005} {\bibfield  {journal} {\bibinfo  {journal} {Phys.
  Rev.}\ }\textbf {\bibinfo {volume} {D84}},\ \bibinfo {pages} {106005}
  (\bibinfo {year} {2011})},\ \Eprint {http://arxiv.org/abs/1106.4804}
  {arXiv:1106.4804 [hep-th]} \BibitemShut {NoStop}%
\bibitem [{\citenamefont {Anderson}\ \emph {et~al.}(2012)\citenamefont
  {Anderson}, \citenamefont {Gray}, \citenamefont {Lukas},\ and\ \citenamefont
  {Palti}}]{Anderson:2012yf}%
  \BibitemOpen
  \bibfield  {author} {\bibinfo {author} {\bibfnamefont {L.~B.}\ \bibnamefont
  {Anderson}}, \bibinfo {author} {\bibfnamefont {J.}~\bibnamefont {Gray}},
  \bibinfo {author} {\bibfnamefont {A.}~\bibnamefont {Lukas}}, \ and\ \bibinfo
  {author} {\bibfnamefont {E.}~\bibnamefont {Palti}},\ }\href {\doibase
  10.1007/JHEP06(2012)113} {\bibfield  {journal} {\bibinfo  {journal} {JHEP}\
  }\textbf {\bibinfo {volume} {06}},\ \bibinfo {pages} {113} (\bibinfo {year}
  {2012})},\ \Eprint {http://arxiv.org/abs/1202.1757} {arXiv:1202.1757
  [hep-th]} \BibitemShut {NoStop}%
\bibitem [{\citenamefont {Nibbelink}, \citenamefont {Loukas},\ and\
  \citenamefont {Ruehle}(2015)}]{Nibbelink:2015vha}%
  \BibitemOpen
  \bibfield  {author} {\bibinfo {author} {\bibfnamefont {S.~G.}\ \bibnamefont
  {Nibbelink}}, \bibinfo {author} {\bibfnamefont {O.}~\bibnamefont {Loukas}}, \
  and\ \bibinfo {author} {\bibfnamefont {F.}~\bibnamefont {Ruehle}},\ }\href
  {\doibase 10.1002/prop.201500041} {\bibfield  {journal} {\bibinfo  {journal}
  {Fortsch. Phys.}\ }\textbf {\bibinfo {volume} {63}},\ \bibinfo {pages} {609}
  (\bibinfo {year} {2015})},\ \Eprint {http://arxiv.org/abs/1507.07559}
  {arXiv:1507.07559 [hep-th]} \BibitemShut {NoStop}%
\bibitem [{\citenamefont {Hall}, \citenamefont {Murayama},\ and\ \citenamefont
  {Nomura}(2002)}]{Hall:2001tn}%
  \BibitemOpen
  \bibfield  {author} {\bibinfo {author} {\bibfnamefont {L.~J.}\ \bibnamefont
  {Hall}}, \bibinfo {author} {\bibfnamefont {H.}~\bibnamefont {Murayama}}, \
  and\ \bibinfo {author} {\bibfnamefont {Y.}~\bibnamefont {Nomura}},\ }\href
  {\doibase 10.1016/S0550-3213(02)00816-7} {\bibfield  {journal} {\bibinfo
  {journal} {Nucl. Phys.}\ }\textbf {\bibinfo {volume} {B645}},\ \bibinfo
  {pages} {85} (\bibinfo {year} {2002})},\ \Eprint
  {http://arxiv.org/abs/hep-th/0107245} {arXiv:hep-th/0107245 [hep-th]}
  \BibitemShut {NoStop}%
\bibitem [{\citenamefont {Hebecker}(2004)}]{Hebecker:2003we}%
  \BibitemOpen
  \bibfield  {author} {\bibinfo {author} {\bibfnamefont {A.}~\bibnamefont
  {Hebecker}},\ }\href@noop {} {\bibfield  {journal} {\bibinfo  {journal}
  {JHEP}\ }\textbf {\bibinfo {volume} {01}},\ \bibinfo {pages} {047} (\bibinfo
  {year} {2004})},\ \Eprint {http://arxiv.org/abs/hep-ph/0309313}
  {hep-ph/0309313} \BibitemShut {NoStop}%
\bibitem [{\citenamefont {Hebecker}\ and\ \citenamefont
  {Trapletti}(2005)}]{Hebecker:2004ce}%
  \BibitemOpen
  \bibfield  {author} {\bibinfo {author} {\bibfnamefont {A.}~\bibnamefont
  {Hebecker}}\ and\ \bibinfo {author} {\bibfnamefont {M.}~\bibnamefont
  {Trapletti}},\ }\href@noop {} {\bibfield  {journal} {\bibinfo  {journal}
  {Nucl. Phys.}\ }\textbf {\bibinfo {volume} {B713}},\ \bibinfo {pages} {173}
  (\bibinfo {year} {2005})},\ \Eprint {http://arxiv.org/abs/hep-th/0411131}
  {hep-th/0411131} \BibitemShut {NoStop}%
\bibitem [{\citenamefont {Trapletti}(2006)}]{Trapletti:2006xv}%
  \BibitemOpen
  \bibfield  {author} {\bibinfo {author} {\bibfnamefont {M.}~\bibnamefont
  {Trapletti}},\ }\href {\doibase 10.1142/S0217732306021785} {\bibfield
  {journal} {\bibinfo  {journal} {Mod.Phys. Lett.}\ }\textbf {\bibinfo {volume}
  {A21}},\ \bibinfo {pages} {2251} (\bibinfo {year} {2006})},\ \Eprint
  {http://arxiv.org/abs/hep-th/0611030} {arXiv:hep-th/0611030 [hep-th]}
  \BibitemShut {NoStop}%
\bibitem [{\citenamefont {Donagi}\ and\ \citenamefont
  {Wendland}(2009)}]{Donagi:2008xy}%
  \BibitemOpen
  \bibfield  {author} {\bibinfo {author} {\bibfnamefont {R.}~\bibnamefont
  {Donagi}}\ and\ \bibinfo {author} {\bibfnamefont {K.}~\bibnamefont
  {Wendland}},\ }\href {\doibase 10.1016/j.geomphys.2009.04.004} {\bibfield
  {journal} {\bibinfo  {journal} {J.Geom.Phys.}\ }\textbf {\bibinfo {volume}
  {59}},\ \bibinfo {pages} {942} (\bibinfo {year} {2009})},\ \Eprint
  {http://arxiv.org/abs/0809.0330} {arXiv:0809.0330 [hep-th]} \BibitemShut
  {NoStop}%
\bibitem [{\citenamefont {Fischer}\ \emph {et~al.}(2013)\citenamefont
  {Fischer}, \citenamefont {Ratz}, \citenamefont {Torrado},\ and\ \citenamefont
  {Vaudrevange}}]{Fischer:2012qj}%
  \BibitemOpen
  \bibfield  {author} {\bibinfo {author} {\bibfnamefont {M.}~\bibnamefont
  {Fischer}}, \bibinfo {author} {\bibfnamefont {M.}~\bibnamefont {Ratz}},
  \bibinfo {author} {\bibfnamefont {J.}~\bibnamefont {Torrado}}, \ and\
  \bibinfo {author} {\bibfnamefont {P.~K.}\ \bibnamefont {Vaudrevange}},\
  }\href {\doibase 10.1007/JHEP01(2013)084} {\bibfield  {journal} {\bibinfo
  {journal} {JHEP}\ }\textbf {\bibinfo {volume} {1301}},\ \bibinfo {pages}
  {084} (\bibinfo {year} {2013})},\ \Eprint {http://arxiv.org/abs/1209.3906}
  {arXiv:1209.3906 [hep-th]} \BibitemShut {NoStop}%
\bibitem [{\citenamefont {Buchm{\"u}ller}\ \emph {et~al.}(2005)\citenamefont
  {Buchm{\"u}ller}, \citenamefont {Hamaguchi}, \citenamefont {Lebedev},\ and\
  \citenamefont {Ratz}}]{Buchmuller:2005sh}%
  \BibitemOpen
  \bibfield  {author} {\bibinfo {author} {\bibfnamefont {W.}~\bibnamefont
  {Buchm{\"u}ller}}, \bibinfo {author} {\bibfnamefont {K.}~\bibnamefont
  {Hamaguchi}}, \bibinfo {author} {\bibfnamefont {O.}~\bibnamefont {Lebedev}},
  \ and\ \bibinfo {author} {\bibfnamefont {M.}~\bibnamefont {Ratz}},\ }in\
  \href@noop {} {\emph {\bibinfo {booktitle} {{Symposium GustavoFest}}}}\
  (\bibinfo {year} {2005})\ pp.\ \bibinfo {pages} {143--156},\ \Eprint
  {http://arxiv.org/abs/hep-ph/0512326} {arXiv:hep-ph/0512326 [hep-ph]}
  \BibitemShut {NoStop}%
\bibitem [{\citenamefont {Blaszczyk}\ \emph {et~al.}(2010)\citenamefont
  {Blaszczyk}, \citenamefont {{Groot Nibbelink}}, \citenamefont {Ratz},
  \citenamefont {Ruehle}, \citenamefont {Trapletti} \emph
  {et~al.}}]{Blaszczyk:2009in}%
  \BibitemOpen
  \bibfield  {author} {\bibinfo {author} {\bibfnamefont {M.}~\bibnamefont
  {Blaszczyk}}, \bibinfo {author} {\bibfnamefont {S.}~\bibnamefont {{Groot
  Nibbelink}}}, \bibinfo {author} {\bibfnamefont {M.}~\bibnamefont {Ratz}},
  \bibinfo {author} {\bibfnamefont {F.}~\bibnamefont {Ruehle}}, \bibinfo
  {author} {\bibfnamefont {M.}~\bibnamefont {Trapletti}},  \emph {et~al.},\
  }\href {\doibase 10.1016/j.physletb.2009.12.036} {\bibfield  {journal}
  {\bibinfo  {journal} {Phys. Lett.}\ }\textbf {\bibinfo {volume} {B683}},\
  \bibinfo {pages} {340} (\bibinfo {year} {2010})},\ \Eprint
  {http://arxiv.org/abs/0911.4905} {arXiv:0911.4905 [hep-th]} \BibitemShut
  {NoStop}%
\bibitem [{\citenamefont {M{\"u}tter}, \citenamefont {Ratz},\ and\
  \citenamefont {Vaudrevange}(2016)}]{Muetter:2016pr}%
  \BibitemOpen
  \bibfield  {author} {\bibinfo {author} {\bibfnamefont {A.}~\bibnamefont
  {M{\"u}tter}}, \bibinfo {author} {\bibfnamefont {M.}~\bibnamefont {Ratz}}, \
  and\ \bibinfo {author} {\bibfnamefont {P.~K.}\ \bibnamefont {Vaudrevange}},\
  }\href@noop {} {\  (\bibinfo {year} {2016})},\ \bibinfo {note} {in
  preparation}\BibitemShut {NoStop}%
\end{thebibliography}%

\end{document}